\documentclass{article}
\usepackage{spconf,amsmath,graphicx,hyperref}
\usepackage{url}            
\usepackage{booktabs}       
\usepackage{nicefrac}       
\usepackage{microtype}      
\usepackage{multirow}
\usepackage{graphicx}
\usepackage{subcaption}
\usepackage{makecell}
\usepackage{setspace}
\usepackage{spconf,amsmath,graphicx,spconf,epsfig,stfloats,hyperref,multirow,booktabs,caption,color,cite,amssymb}

\title{DECAF: A Privacy Preserving Speech Codec Using Speaker Disentanglement and Canonical Voice Conversion}
\name{\parbox{\linewidth}{\centering
    Md Shakhrul Iman Siam\sthanks{Shakhrul was an intern at Microsoft during the course of this work.}$^1$, Dushyant Sharma$^2$, Stanislav Yu. Kruchinin$^2$ and Peter Skala$^2$
}}

\address{
    $^1$Ohio State University, USA\\
    $^2$Microsoft Health \& Life Sciences AI
}

\begin{document}
%
\maketitle
\begin{abstract}
We present DECAF, a privacy-preserving neural speech codec that obfuscates a speaker's voice while preserving linguistic content while maintaining automatic speech recognition (ASR) performance at very low bitrates, inspired by decaffeination. At the transmitter end, speech is encoded into speaker-independent content embeddings, which are compressed using residual vector quantization and transmitted without any speaker-related information. At the receiver, a canonical speaker embedding, shared a priori between endpoints, is used for waveform reconstruction, enabling deterministic and consistent obfuscation of a speaker's voice. The proposed framework leverages an information bottleneck applied to self-supervised representations, along with a separate speaker embedding branch, to achieve effective speaker–content disentanglement. We further incorporate a CTC-based auxiliary objective, encouraging content representations that are well aligned with downstream ASR tasks. We show that DECAF operating at a bit rate of 0.5~kbps achieves an Equal Error Rate~(EER) of up to 43.5\% for a speaker verification system, while maintaining competitive ASR performance, yielding a relative reduction in word error rate of 33.2\% compared to a state of the art method. 

\end{abstract}

\section{Introduction}
\label{sec:intro}
As voice enabled interfaces become ubiquitous, greater emphasis must be placed on the need to improve privacy through speaker obfuscation. A user's voice data is often transmitted to the cloud for computational reasons and then subsequently stored for model improvements. In a speech signal, identity can be revealed through the content (e.g. mention of a name) or through the speaker’s voice itself. The problem of effectively concealing a speaker's voice in a speech signal while preserving intelligibility is the focus of this work. Existing approaches for speaker obfuscation falls into three categories. \textit{Signal-processing based methods} manipulate acoustic features without training, e.g., vocal-tract-length normalization~\cite{qian2017voicemask} or the McAdams transformation~\cite{patino2020speaker}. \textit{Voice-conversion (VC) based methods} replace source-speaker characteristics with those of a target while preserving content, including x-vector-based anonymization~\cite{fang2019speaker}, adversarial disentanglement~\cite{espinoza2020speaker}, and ASR-bottleneck suppression of speaker identity~\cite{srivastava2019privacy}. \textit{Adversarial-perturbation methods} inject inconspicuous distortions to fool automatic speaker verification~(ASV) systems: V-Cloak~\cite{deng2023v} modulates Wave-U-Net features per frequency band, and VoiceBlock~\cite{o2022voiceblock} applies a time-varying FIR filter. Recent work on Neural Audio CODEC~(NAC) approaches~\cite{panariello2024speaker} and canonical-VC~\cite{sharma2023canonical} extend these ideas. These approaches share a structural assumption: obfuscation and transmission are separate problems. A waveform is first obfuscated, then encoded by a separate codec, so the bandwidth cost is paid twice and the system inherits the privacy/utility trade-off of its obfuscation stage. The current NAC-based state of the art~\cite{panariello2024speaker} method degrades ASR performance by several percent absolute WER, even at high bitrates. There is a large tradeoff in obfuscation and utility.

In this paper, we propose DECAF, a neural voice CODEC that treats compression as part of the obfuscation mechanism. The transmitter applies a Connexionist Temporal Classification~(CTC)~\cite{graves2006ctc} supervised information bottleneck that strips speaker identity from the input while preserving linguistic content, and only the resulting content stream is quantized and transmitted. The receiver reconstructs the waveform using a canonical speaker embedding that is shared a priori between transmitter and receiver. Two properties follow from this design. First, no direct speaker information is transmitted, so the bandwidth cost is paid once rather than twice as in obfuscation-then-encode pipelines. Second, the dimensionality of the bottleneck places an architectural upper bound on the speaker information that can survive transmission, making privacy a property of the codec itself. DECAF jointly satisfies three objectives prior work has typically pursued in isolation: speaker obfuscation robust to an ASV attacker, low-bitrate transmission and preserved ASR utility. In our work we target an ASR application where the main task is speech to text.

Our key contribution is a novel codec pipeline: content-only transmission with canonical speaker reconstruction. DECAF utilizes the FreeVC~\cite{li2023freevc} architecture and extends it by introducing an additional CTC loss in the speaker-content disentanglement stage that produces content embeddings that are more ASR-compatible. Our second contribution is to quantize and transmit only the content embeddings at very low bitrates by utilizing a canonical speaker embedding that is shared a priori. Lastly, we show that the smaller, WavLM-base+ SSL model can be used to obtain better results than the WavLM-Large model, resulting in a 3x speedup in inference speed for WavLM on a V100 GPU. We show that fine-tuning a frozen Whisper-medium recognizer on DECAF-decoded audio at 0.5~kbps achieves a WER that is 33\% lower relative to a state of the art system and within 1.8\% absolute of original-audio WER, with a 43.5\% EER. 
\vspace*{-5pt}
\section{DECAF: Disentangling Encoder for Compressed Audio Features}
\vspace*{-5pt}
\label{sec:method}
\begin{figure*}[t]
    \centering
    \includegraphics[width=0.9\linewidth]{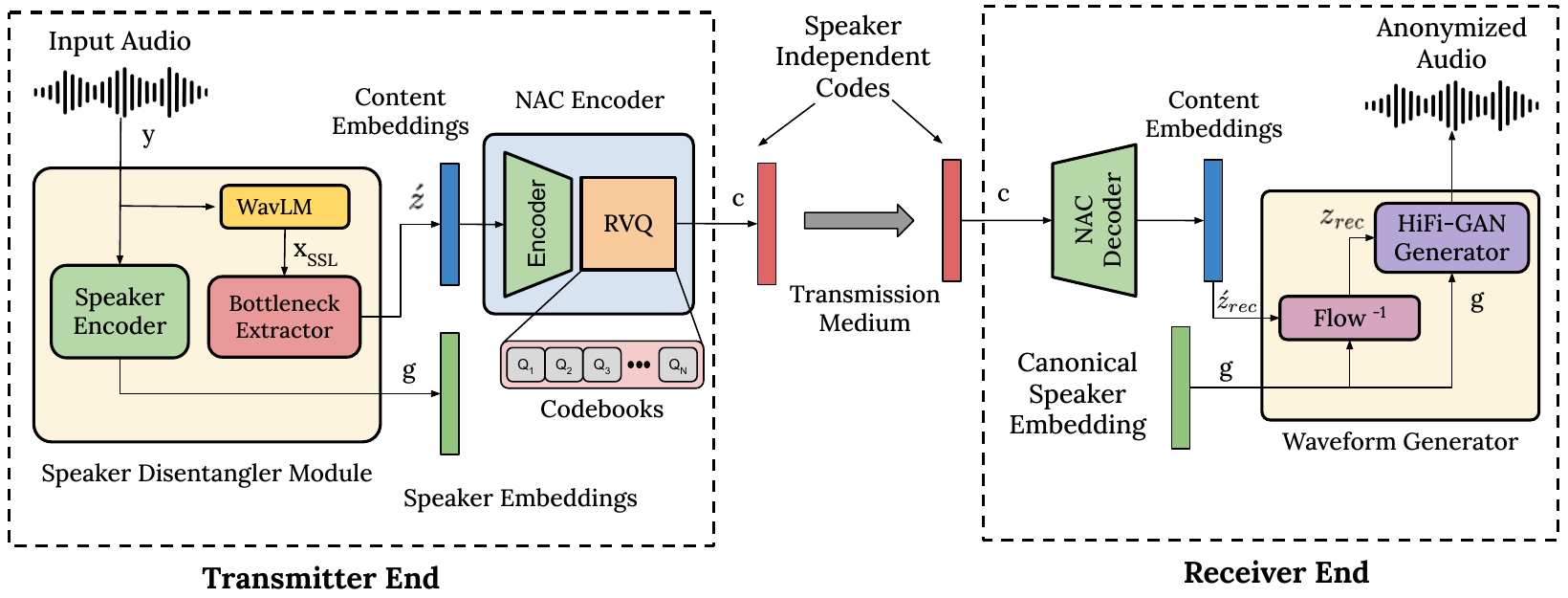}
    \caption{The proposed DECAF system. Since a canonical speaker embedding is used at the receiver end, the speaker encoder can be disabled altogether and a speaker embedding does not need to be transmitted. It is only needed during training.}
    \label{fig:overview}
\end{figure*}

\begin{figure}
    \centering
    \includegraphics[width=\linewidth]{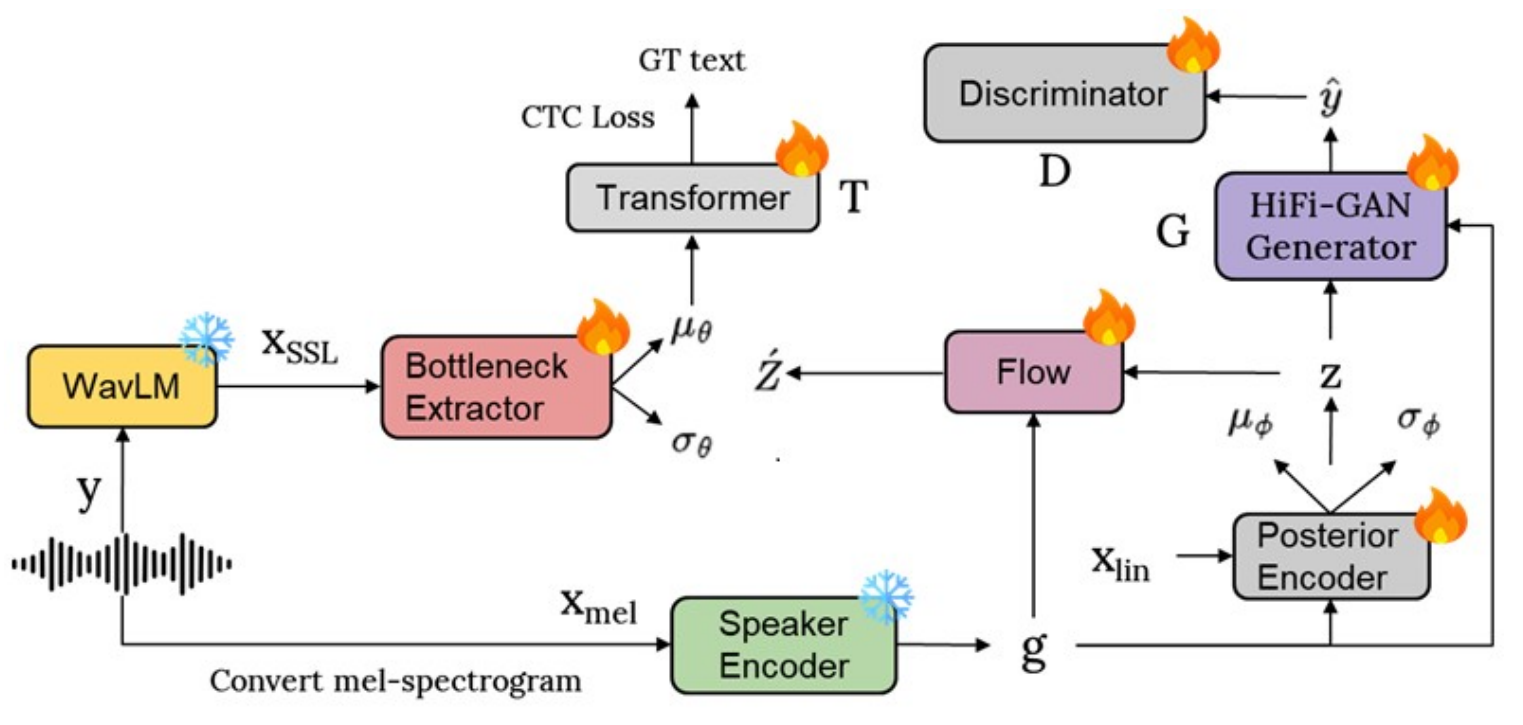}
    \caption{Training Process of Speaker Disentangle module.}
    \label{fig:training freevc}
\end{figure}
DECAF (Fig.~\ref{fig:overview}) consists of a speaker disentangler module~(SDM), an NAC (comprising an encoder and decoder), a waveform generator, and a canonical speaker embedding. 
\subsection{Speaker Disentangler Module~(SDM)}
The Speaker Disentangler Module (SDM) produces features that are simultaneously stripped of speaker identity and rich enough to support automatic speech recognition, even after quantization to very low bitrates. We adopt the disentanglement backbone of FreeVC~\cite{li2023freevc}, which applies an information bottleneck to self-supervised learning~(SSL) features to suppress speaker identity while preserving linguistic content. This includes a WavLM~\cite{chen2022wavlm} frontend that produces high-level acoustic features, a WaveNet-based bottleneck extractor~\cite{van2016wavenet} that compresses these features to a low-dimensional posterior $\mathcal{N}(\acute{z}; \mu_\theta, \sigma_\theta^2)$, and a speaker encoder~\cite{liu2021any}. We refer to the variant using WavLM-base+ as DECAF-small and the variant using WavLM-large as DECAF-large. The dimensionality gap between WavLM's output (1024-dim for WavLM-large and 768-dim for WavLM-base+) and bottleneck extractor (192-dim) induces an information bottleneck, compelling the model to discard content-irrelevant information, such as background noise and speaker information, thereby preserving only the core linguistic content. FreeVC's content embedding is optimized for waveform reconstruction in a voice conversion setup, not for downstream speech recognition. Thus, we augment the FreeVC objective with a CTC loss applied to output of a small transformer encoder network placed onto the content embedding, which forces the features to remain phonetically discriminant through the bottleneck. \\
We train the SDM module on the VCTK~\cite{yamagishi2019vctk} dataset with the additional CTC loss as shown in Fig~\ref{fig:training freevc}. The model is trained with a learning rate of $2\times10^{-4}$ for 450k steps with a batch size of 64. The generator-related loss is defined as \(L_{\text{Gen}}=L_{\text{rec}}+L_{\text{KL}}+L_{\text{adv}}(G)+L_{\text{fm}}(G)+L_{\text{CTC}}\). Here \(L_{\text{rec}}\) is the reconstruction loss computed as the $\text{L}_{1}$ distance between target and predicted mel-spectrogram, and \(L_{\text{KL}}\) is the KL divergence between the prior distribution \(p_{\theta}(\acute{z}\mid x_{\text{ssl}})\) and the posterior distribution \(q_{\theta}(z\mid x_{\text{lin}})\), where $x_{\text{lin}}$ denotes linear spectrogram, $x_{\text{ssl}}$ denotes SSL feature. 
The CTC loss is computed between estimated and true transcriptions by passing the $\acute{z}$ embeddings through a transformer encoder \(T\), comprising 6 layers, 8 attention heads and a feedforward dimension of 2048. \(L_{\text{fm}}(G)\) is the feature matching loss for generator \(G\), and \(L_{\text{adv}}(G)\) is the adversarial loss for generator \(G\). The discriminator is trained using an adversarial loss \(L_{\text{Dis}}=L_{\text{adv}}(D)\).

\subsection{Neural Audio CODEC~(NAC)}
The NAC is an Encoder-RVQ-Decoder structure which compresses the content embeddings ($\acute{z}$). The encoder and decoder use linear layers for matching input dimensions to the latent dimension. We use Residual Vector Quantization~(RVQ) to quantize the output of the encoder into a set of codes. RVQ refines the quantization process by computing the residual error after each quantization step and further quantizing it using a codebook of size 1024, with the number of quantizers varied in different bitrate setups. The NAC is trained on the 960~hr training partition of LibriSpeech for 20 epochs with a learning rate of $10^{-4}$ and batch size of 64. 
During training, the input embeddings $\acute{z}$ are produced by the frozen SDM module trained in the previous stage.
The training loss is composed of a reconstruction loss and an RVQ commitment loss \cite{defossez2022highfi}. 
The reconstruction loss is computed by minimizing the $\text{L}_{2}$ distance between the input embeddings $\acute{z}$ and reconstructed embeddings $\acute{z}_{rec}$ as, $L_{\text{reconstruction}} = \left\| \acute{z} - \acute{z}_{rec} \right\|_2^2  $.
\noindent
For each residual step $c \in \{1, ... C\}$, denoting $z_c$ as the current residual and $q_c(z_c)$ as the nearest entry in the corresponding codebook, the RVQ commitment loss is defined as~\eqref{eq:comitloss}.
\begin{equation}
\label{eq:comitloss}
    L_{\text{commitment}}  = \sum_{c=1}^{C} \left\| z_c - q_c(z_c) \right\|_2^2
\end{equation}

\noindent
The total loss of the NAC ($L_{\text{NAC}}$) is computed as $L_{\text{NAC}} = L_{\text{reconstruction}} + 0.25 \times L_{\text{commitment}}$.


\subsection{Waveform Generator}
The waveform generator reconstructs audio from the decoded content and speaker embeddings. It consists of a normalizing flow network \cite{dinh2016density} followed by a HiFi-GAN decoder\cite{kong2020hifi}. 
The waveform generator module is trained jointly with the SDM module as shown in Fig~\ref{fig:training freevc}. We use a single, canonical speaker embedding taken from the Librispeech training set to reconstruct all audio, thus achieving speaker obfuscation. 



\section{Evaluation}
\begin{table*}[t]
\centering
\caption{Results on LS test sets with original and reconstructed audio input. WER finetuned is for the Whisper-medium fine tuned on the DECAF-small CODEC operating at 0.5~kbps. The DECAF systems marked by * are without the NAC component. }
\label{tab:table-1}
\resizebox{\textwidth}{!}{
\renewcommand{\arraystretch}{0.9}
\begin{tabular}{l c | ccc | ccc | ccc | ccc}
\toprule
\multirow{2}{*}{\textbf{Input Audio}} & \multirow{2}{*}{\textbf{Bitrate}} &
\multicolumn{3}{c}{\textbf{EER} (\%) $\uparrow$} &
\multicolumn{3}{c}{\textbf{WER} (\%) $\downarrow$} &
\multicolumn{3}{c}{\textbf{WER-FT} (\%) $\downarrow$} & \multicolumn{3}{c}{\textbf{Speech Quality}}\\
\cmidrule(lr){3-5} \cmidrule(lr){6-8} \cmidrule(lr){9-11} \cmidrule(lr){12-14}
 & (kbps) & \textbf{clean} & \textbf{other} & \textbf{mean} & \textbf{clean} & \textbf{other} & \textbf{mean} & \textbf{clean} & \textbf{other} & \textbf{mean}  & \textbf{PESQ}& \textbf{STOI}&\textbf{XANE}\\
\midrule
Original Audio & 256 & 1.62 & 1.56 & 1.59 & 2.88 & 8.20 & 5.54 & 2.03&  5.25&  3.64& -& -&-\\
FreeVC & 256 & 36.13 & 23.74 & 29.94 & 3.29 & 8.47 & 5.88 & -& -&  -& -& -&-\\
Spk Anon (random)& 256 & 45.62 & 42.34 & 43.98 & 7.84 & 20.23 & 14.04 & -& -&  -& -& -&-\\
Spk Anon (canonical)& 256& 46.96& \textbf{43.99}& \textbf{45.47}& 7.44& 17.50& 12.47& 4.05& 12.38& 8.22& -& -&-\\
DECAF-small* & 153 & 42.63 & 30.32 & 36.48 & 3.34 & 8.69 & 6.02 & 2.25& 5.84&  4.05& -& -&-\\
DECAF-large* & 153 & 36.28 & 24.96 & 30.62 & 3.32 & 8.49 & 5.91 & 2.46& 6.73&  4.60& -& -&-\\
\midrule
\midrule
Encodec & 1.5 & 24.67 & 22.18 & 23.43 & 6.41 & 21.44 & 13.93 & -& -&  -& 1.57& 0.84&0.96\\
Encodec & 3.0 & 21.46 & 19.49 & 20.48 & 3.35 & 11.90 & 7.63 & -& -&  -& 2.09& 0.89&0.97\\
Encodec & 6.0 & 19.12 & 17.65 & 18.39 & 3.09 & 9.45 & 6.27 & -& -&  -& 2.73& 0.93&0.97\\
Encodec & 12.0 & 17.88 & 16.19 & 17.04 & 3.01 & 8.43 & 5.72 & -& -&  -& 3.31& 0.96&0.97\\
Encodec & 24.0 & 17.24 & 15.55 & 16.40 & 3.01 & 7.97 & 5.49 & -& -&  -& 3.63& 0.97&0.97\\
\midrule
DECAF-small & \textbf{0.5} & \textbf{47.59} & 39.36 & 43.48 & 5.46 & 12.57 & 9.02 & 2.85 & 8.13 & 5.49  & 1.58& 0.89&0.95\\
DECAF-small & 1.0 & 45.31 & 35.12 & 40.22 & 4.19 & 10.55 & 7.37 & 2.66 & 6.81 & 4.74  & 2.19& 0.94&0.95\\
DECAF-small & 1.5 & 44.08 & 32.64 & 38.36 & 3.49 & 9.69 & 6.59 & 2.46 & 6.35 & 4.41  & 2.81& 0.96&0.95\\
DECAF-small & 3.0 & 43.04 & 31.23 & 37.14 & 3.40 & 8.40 & 5.90 & 2.24 & 6.05 & 4.15  & 3.60& 0.98&0.95\\
DECAF-small & 6.0 & 42.62 & 30.58 & 36.60 & 3.34 & 9.12 & 6.23 & 2.34 & 6.05 & 4.20  & 4.15& 0.99&0.95\\
DECAF-small & 12.0 & 42.62 & 30.37 & 36.50 & 3.35 & 8.91 & 6.13 & 2.31 & 6.06 & 4.19  & 4.48& \textbf{1.0}&0.95\\
DECAF-small & 24.0 & 42.59 & 30.35 & 36.47 & 3.35 & 8.85 & 6.10 & 2.38 & 6.11 & 4.25  & \textbf{4.57}& \textbf{1.0}&0.95\\
\midrule
DECAF-large & 0.5 & 43.17 & 33.74 & 38.46 & 5.50 & 15.20 & 10.35 & 3.00 & 9.32 & 6.16  & 1.79& 0.91&0.94\\
DECAF-large & 1.0 & 38.85 & 27.73 & 33.29 & 3.72 & 9.91 & 6.82 & 2.48 & 6.60 & 4.54  & 2.59& 0.95&0.94\\
DECAF-large & 1.5 & 37.80 & 26.61 & 32.21 & 3.42 & 9.16 & 6.29 & 2.46 & 6.36 & 4.41  & 3.18& 0.97&0.95\\
DECAF-large & 3.0 & 36.70 & 25.50 & 31.10 & 3.36 & 8.69 & 6.03 & 2.38 & 6.22 & 4.30  & 3.69& 0.98&0.95\\
DECAF-large & 6.0 & 36.41 & 25.14 & 30.78 & 3.30 & 8.49 & 5.90 & 2.39 & 6.17 & 4.28  & 4.16& 0.99&0.95\\
DECAF-large & 12.0 & 36.24 & 25.00 & 30.62 & 3.38 & 8.31 & 5.85 & 2.36 & 6.23 & 4.30  & 4.44& \textbf{1.0}&0.95\\
DECAF-large & 24.0 & 36.29 & 24.91 & 30.60 & 3.33 & 8.44 & 5.89 & 2.35 & 6.24 & 4.30  & \textbf{4.57}& \textbf{1.0}&0.95\\
\bottomrule
\end{tabular}
}
\end{table*}

\subsection{Data}
The training of the SDM and NAC components is based on the VCTK~\cite{yamagishi2019vctk} and Librispeech~(LS) 960~hr~\cite{Panayotov2015-LAA} datasets, respectively. In order to evaluate the ASR performance of the various systems, we use the Whisper-medium multilingual ASR~\cite{radford2022whisper} model and also fine-tune it on LS training data processed through DECAF at various bit rates. All testing is done on the test-clean and test-other partitions of LS.
\vspace*{-5pt}
\subsection{Evaluation Metrics}
\noindent
\textbf{Equal Error Rate (EER):}
A standard metric used to evaluate speaker verification systems. We use an ECAPA-TDNN based ASV model~\cite{huggingface2025}.\\
\noindent
\textbf{Word Error Rate (WER):} We use the Whisper-medium multilingual ASR~\cite{radford2022whisper} model with its corresponding text normalizer and the evaluation package from~\cite{huggingface2025}.\\
\noindent
\textbf{Perceptual Evaluation of Speech Quality (PESQ):} An intrusive metric correlated with subjective audio quality~\cite{recommendation2001perceptual,torchmetrics}.\\
\noindent
\textbf{Short-Time Objective Intelligibility (STOI):} An intrusive metric correlated with speech intelligibility scores from listening tests~\cite{stoi,pystoi}.\\
\noindent
\textbf{Cosine Similarity of XANE embeddings:}
XANE~\cite{dumpala2024xane,sharma2024xane} extracts neural embeddings that model the background acoustics of a speech signal. We compute the cosine similarity between XANE embeddings from the original and processed audio.
\vspace*{-5pt}
\subsection{Baselines}
\vspace*{-3pt}
\noindent
DECAF targets an operating point that no single prior system addresses: simultaneous extreme compression, speaker obfuscation, and preserved ASR utility. We therefore compare against baselines that each cover a subset of these objectives.
Encodec is a state-of-the-art NAC~\cite{defossez2022highfi} that employs a convolutional encoder-decoder architecture with an RVQ and allows operating at very low bit rates. We also use a current state of the art, NAC-based speaker anonymization system (denoted as Spk Anon)~\cite{panariello2024speaker}, configured in the canonical voice conversion mode as well as the random mode.
\section{Experiments}
We evaluate the DECAF system and the baseline methods using the LibriSpeech test sets and present the summarized results in Table~\ref{tab:table-1}. The speaker anonymization performance is measured using the EER metric and both DECAF models (base and large) outperform the Encodec baselines. The Spk Anon system has a very good performance for speaker anonymization but with a very large degradation in ASR performance (nearly 9\% WER higher absolute than the original audio). Notably, across all models, a reduction in bitrate consistently leads to an increase in EER, suggesting that lower bitrates enhance anonymization.

The DECAF models demonstrate competitive WER performance relative to both the original audio and the baseline models, indicating a favorable trade-off between speaker anonymization and intelligibility. The ASR accuracy is further improved with fine-tuning the whisper model on DECAF-small 0.5~kbps processed data. The results, shown in the WER-FT column, reveal a substantial reduction in WER post fine-tuning, surpassing all baseline models and even the original audio based model on original audio by nearly 1\% relative. These findings underscore that DECAF, when fine-tuned appropriately, can achieve robust ASR performance while ensuring secure transmission of voice data through effective speaker anonymization. To assess the impact of the NAC on speech quality and intelligibility, we evaluate DECAF using three metrics, as presented in the last columns of Table~\ref{tab:table-1}. For PESQ and STOI metrics, the transmitted audio from the Encodec and DECAF models serves as the reference, while the received (decoded) audio is treated as the target. The results demonstrate that both DECAF-small and DECAF-large outperform the baseline models across all bitrate settings. At higher bitrates (e.g.~24~kbps), DECAF achieves near-transparent reconstruction, with both PESQ and STOI scores reaching their respective maximum values, indicating minimal perceptual and intelligibility loss during transmission. Even at significantly lower bitrates (e.g. 1.5 kbps), DECAF maintains superior performance, whereas the baseline models exhibit substantial degradation in both quality and intelligibility. 
To access the impact on background acoustics, we measure XANE cosine similarity. Higher values indicates maximum background acoustics retention. Both DECAF-small and DECAF-large retain less background acoustic information across all tested bitrates. Lower values here indicate that less of the background information is preserved, which is an additional benefit for voice privacy.


\section{Conclusions}

We presented DECAF, a neural voice CODEC that performs speaker voice obfuscation by disentangling, compressing and transmitting only the content embeddings. At the transmitter end our method strips the speaker's voice from the input audio a new CTC-supervised information bottleneck and transmits only the quantized content stream. At the receiver end, the codec reconstructs the waveform using a canonical speaker embedding that is shared apriori between endpoints. Operating at a bit rate of just 0.5~kbps, DECAF achieves an EER of 43.5\% against an ECAPA-TDNN speaker verification attacker while keeping the WER within 1.8\% absolute of the original audio with the Whisper-medium ASR model after fine-tuning. If used without any compression, the DECAF WER is only 0.44\% higher than original audio based ASR but still with a nearly 35 times higher EER (i.e. similar utility with much higher privacy). In contrast, a NAC based state of the art speaker obfuscation method achieves an average EER of 45.5\% but the WER increases by 4.6\% relative to the original audio, representing a nearly 75\% relative WER increase compared with DECAF at 0.5~kbs. 



\begingroup
\setstretch{0.7}
\bibliographystyle{IEEEbib}
\bibliography{refs_p1}
\endgroup


\end{document}